\documentclass[conference]{IEEEtran}
\IEEEoverridecommandlockouts
\usepackage{cite}
\usepackage{amsmath,amssymb,amsfonts}
\usepackage{algorithmic}
\usepackage{graphicx}
\usepackage{textcomp}
\usepackage{xcolor}
\usepackage{url}

\usepackage{tikz}
\usetikzlibrary{graphs, positioning, quotes, shapes,arrows}

\usepackage{selinput}\SelectInputMappings{adieresis={ä},germandbls={ß}}
\usepackage{algorithmic}
\usepackage[ruled,vlined,linesnumbered,commentsnumbered]{algorithm2e}
\usepackage{leftidx}
\usepackage{amsthm}
\usepackage[capitalise]{cleveref}
\usepackage{soul}
\usepackage{svg}
\usepackage{subfigure}
\usepackage{array}
\usepackage{multirow}
\usepackage{hhline}
\usepackage{booktabs}
\usepackage[font=footnotesize]{caption}
\usepackage{array}
\newcolumntype{L}[1]{>{\raggedright\let\newline\\\arraybackslash\hspace{0pt}}m{#1}}
\newcolumntype{C}[1]{>{\centering\let\newline\\\arraybackslash\hspace{0pt}}m{#1}}
\newcolumntype{R}[1]{>{\raggedleft\let\newline\\\arraybackslash\hspace{0pt}}m{#1}}

\def\BibTeX{{\rm B\kern-.05em{\sc i\kern-.025em b}\kern-.08em
    T\kern-.1667em\lower.7ex\hbox{E}\kern-.125emX}}

\usepackage[shortcuts,acronym]{glossaries}
\newacronym{aeica}{AeICA}{Adaptive Extraction-Based Independent Component Analysis}
\newacronym{auc}{AUC}{Area Under the receiver operating characteristic Curve}
\newacronym{ai}{AI}{Artificial Intelligence}

\newacronym{bss}{BSS}{Blind Source Separation}

\newacronym{cia}{C-IA-Net}{Comparison IA-Net}
\newacronym{cots}{COTS}{Commercial off-the-shelf}
\newacronym{coin}{COIN}{Computing in the Network}
\newacronym{comnetsemu}{ComNetsEmu}{Communication Networks Emulator}
\newacronym{cf}{C-F}{Compute-and-Forward}
\newacronym{cps}{CPS}{Cyber-Physical System}

\newacronym{ica}{ICA}{Independent Component Analysis}
\newacronym{iot}{IoT}{Internet of Things}
\newacronym{iiot}{IIoT}{Industrial Internet of Things}
\newacronym{ia-net}{IA-Net}{Information Abstraction Neural Network}
\newacronym{ia-net-lite}{IA-Net-Lite}{Information Abstraction Neural Network Lite}

\newacronym{mauc}{mAUC}{mean-AUC}
\newacronym{mac}{MAdd}{Multiply-Add Operation}
\newacronym{mtu}{MTU}{Maximum Transmission Unit}

\newacronym{nf}{NF}{Network Function}
\newacronym{nfv}{NFV}{Network Functions Virtualization}
\newacronym{nn}{DNN}{Deep Neural Network}

\newacronym{pauc}{pAUC}{partial-AUC}

\newacronym{ria}{R-IA-Net}{Reference IA-Net}
\newacronym{roc}{ROC}{Receiver Operating Characteristic}
\newacronym{rtt}{RTT}{Round Trip Time}

\newacronym{sdn}{SDN}{Software Defined Networking}
\newacronym{sfc}{SFC}{Service Function Chain}
\newacronym{sf}{S-F}{Store-and-Forward}

\newacronym{ts}{TS Method}{two-stage method}

\newacronym{udp}{UDP}{User Datagram Protocol}

\newacronym{vnf}{VNF}{Virtual Network Function}

\begin{document}
\title{In-Network Processing for Low-Latency Industrial Anomaly Detection in Softwarized Networks}
\author{
    \IEEEauthorblockN{
        Huanzhuo Wu\IEEEauthorrefmark{1},
        Jia He\IEEEauthorrefmark{1},
        Máté Tömösközi\IEEEauthorrefmark{1}\IEEEauthorrefmark{2},
        Zuo Xiang\IEEEauthorrefmark{1} and
        Frank H.P. Fitzek\IEEEauthorrefmark{1}\IEEEauthorrefmark{2}\\
    }
    \IEEEauthorblockA{\IEEEauthorrefmark{1}Deutsche Telekom Chair of Communication Networks - Technische Universität Dresden, Germany
    }
    \IEEEauthorblockA{\IEEEauthorrefmark{2}Centre for Tactile Internet with Human-in-the-Loop (CeTI)}
    \emph{Email:\{huanzhuo.wu$\mid$mate.tomoskozi$\mid$zuo.xiang$\mid$frank.fitzek\}@tu-dresden.de, jia.he@mailbox.tu-dresden.de}
    \thanks{This is a preprint of the work~\cite{Wu21:InNetworkProcessingLowLatency}, that has been accepted for publication in the proceedings of the 2021 IEEE Global Communications Conference.}
}

\maketitle

\begin{abstract}
Modern manufacturers are currently undertaking the integration of novel digital technologies -- such as 5G-based wireless networks, the Internet of Things (IoT), and cloud computing -- to elevate their production process to a brand new level, the level of smart factories.
In the setting of a modern smart factory, time-critical applications are increasingly important to facilitate efficient and safe production.
However, these applications suffer from delays in data transmission and processing due to the high density of wireless sensors and the large volumes of data that they generate.
As the advent of next-generation networks has made network nodes intelligent and capable of handling multiple network functions, the increased computational power of the nodes makes it possible to offload some of the computational overhead.
In this paper, we show for the first time our IA-Net-Lite industrial anomaly detection system with the novel capability of in-network data processing.
IA-Net-Lite utilizes intelligent network devices to combine data transmission and processing, as well as to progressively filter redundant data in order to optimize service latency.
By testing in a practical network emulator, we showed that the proposed approach can reduce the service latency by up to 40\%.
Moreover, the benefits of our approach could potentially be exploited in other large-volume and artificial intelligence applications.


\end{abstract}


\section{Introduction}\label{sec:introduction}
Presently, the general industrial community is facing the challenge of integrating Industry 4.0 technologies into their manufacturing processes, which will eventually lead towards full digitization and intelligence empowered by such emerging technologies as \ac{iot}, artificial intelligence (AI), and cloud computing~\cite{wollschlaeger2017future}.
Among these, \ac{iot} devices are expected to act as both data collectors and the actuators of remote intelligent computing agents, thereby closing the full control loop in order to realize a fully autonomous factory operation.
In future smart factories, \ac{iot} devices would be tasked with the monitoring of processes and machines, etc.
The data produced by such \ac{iot} devices can be, in turn, employed to detect potential disturbances in the production process before an error happens (\emph{predictive maintenance}), which can decrease production downtime and maintenance costs of smart factories~\cite{civerchia2017industrial}.

In particular, time-critical acoustic anomaly detection in industrial environments carries great potential, since acoustic data is able to foreshadow potential failures in the early-stage and offers significant gains to accurately predict when maintenance work is crucial~\cite{10.1145/1541880.1541882}.
Unlike traditional industrial signals -- e.g., temperature and humidity -- acoustic signals are generally distorted before an abnormal event occurs. Also, such audio-based detection methods can avoid blocking and angular distortion problems compared to the alternative image analysis techniques, as well.

\ac{iot}-enabled acoustic sensors deployed in factory halls normally collect and transmit the raw sensor readings for further analysis to separate dedicated devices.
In a generic setup, several devices are deployed at the production site, which creates a multi-object acoustic anomaly detection environment.
Particularly, the observed raw sensory acoustic signal is inevitably altered (mixed) due to the presence and operation of multiple objects.
Depending on the complexity of the deployed sensors and observation systems, the overall amount of data generated can easily reach several Gbps ranges, which can be understood as a big data problem with \emph{data-rich, information-poor} characteristics~\cite{8291112}.
In~\cite{Wu2106:Abstraction}, a data process based on a \ac{nn} and \ac{ia-net} is proposed to address the challenge of \emph{data-rich, information-poor} environments, which empowers detection techniques with high-information density (the acoustic deviation).
After \ac{ia-net} solved the issue of processing large amounts of \ac{iot} data for anomaly detection, we are now facing the question of \emph{where should \ac{ia-net} do the processing}?


The most straightforward solution is to employ a centralized service that manages the handling of the data stemming from the growing volume of sensors. This effectiveness relies on the local Internet access, which is often limited to a few hundred Mbps.
While large companies can simply upgrade their infrastructure, most small and mid-sized industrial manufacturers rely on traditional ISPs for their networking needs, and higher access speeds are tied to higher costs~\cite{glebke2019case}.
A different approach would be to deploy \ac{ia-net} on an edge cloud physically close to the targeted machines.
Compared with the centralized approach, the edge cloud has an advantage in data transmission time, however, its resource-constrained characteristics will lead to more delay in data processing and, ultimately, would prevent the fulfillment of time-critical services.

With the aid of network softwarization, it is possible to deploy flexible services and applications by combining local, edge, and remote computing resources within the same logical network~\cite{glebke2019towards}.
Typically, the deployment of softwarized networks includes a combination of technologies: \ac{sdn}~\cite{haleplidis2015software}, \ac{nfv}~\cite{han2015network}, and \ac{sfc}~\cite{halpern2015service}. 
Utilizing available computational resources within the network for in-network processing concepts, computation steps can be deployed in the on-premise network to reduce the overall data volume and, potentially, speed up processing times~\cite{irtf-coinrg-use-cases-00}.
Therefore, a major challenge is to filter out redundant data by applying the processing logic of information abstraction on resource-constrained network nodes.

In this paper, based on the idea of information abstraction proposed in~\cite{Wu2106:Abstraction}, we show the novel design of a lightweight \ac{nn} model called \ac{ia-net-lite}, and propose its deployment for the first time in softwarized networks, which, contrary to present norms, enables in-network data processing to achieve low-latency industrial anomaly detection.
Extensive evaluations of practical implementations show that \ac{ia-net-lite} reduces the computation load by up to a quarter of the original and decreases the data transfer delays by about $34.38\%$, thus achieving an overall latency reduction of up to $38.89\%$.


The rest of the paper is structured as follows.
First, \cref{sec:background} introduces the idea of information abstraction and describes the problem, after which the proposed method is described in~\cref{sec:method}.
In~\cref{sec:experimental_design}, we introduce the experimental setup and its metrics.
\cref{sec:results_analysis} covers the numerical evaluation and a discussion about our measurement results.
Finally, \cref{sec:conclusion} concludes our contribution and points out future research aspects.
\section{Background}\label{sec:background}


In this section, we first describe our system for acoustic anomaly detection and the methods used to filter out redundant information.
Following that, we discuss the importance of this redundant information filtering approach for low-latency services and the challenges when exploiting it in networks.

\subsection{Information Abstraction}

\begin{figure}[t!]
\centerline{\includegraphics[width=.6\columnwidth]{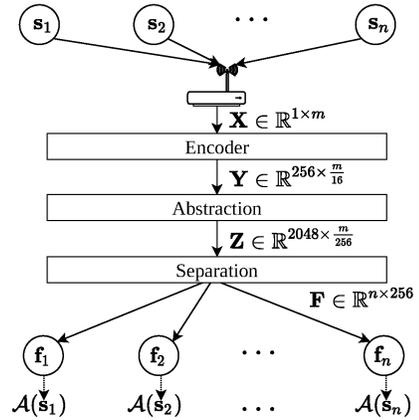}}
\caption{The schematics of information abstraction for industrial anomaly detection.}
\label{fig:setup_ads}
\end{figure}

In multi-machine environments, such as factories, a number of source machines ($n$) are expected to be detected as source objects, see~\cref{fig:setup_ads}. The source of acoustic data is generated in $m$ time slots and is denoted as $\{\mathbf{s}_1, \mathbf{s}_2, \cdots, \mathbf{s}_n\} = \mathbf{S} \in \mathbb{R}^{n \times m}$.
Due to the presence and operation of multiple objects, the observation $\mathbf{X} \in \mathbb{R}^{1 \times m}$ of \ac{iot} sensors is inevitably a mixture of the $n$ source data.
$\mathbf{X}$, in turn, is transmitted to the data processing system to determine irregularities (or anomalies) related to the $n$ objects when only $\mathbf{X}$ is known.

Such an anomaly is defined by the anomaly score and is denoted as $\mathcal{A}$.
$\mathcal{A}$ represents the difference between the actual data and a reference $\mathbf{S}^r$. 
When $\mathcal{A}$ exceeds a threshold value, the corresponding object is declared to be an anomaly, as shown in~\cref{fig:setup_ads}.
In order to get $\mathcal{A}$, more attention needs to be paid to the deviation between the detected and the reference objects than to the detected object itself.
Therefore, redundant information not related to exceptions should be eliminated and only exception-related information needs to be extracted.
Based on this intuition, we proposed in~\cite{Wu2106:Abstraction} the idea of information abstraction to abstract separated anomalous information from raw sensory data for multi-object acoustic anomaly detection, which is a particularly advantageous data processing approach when handling large data volumes.

As shown in \cref{fig:setup_ads}, the information abstraction consists of three distinct modules: encoder, abstraction, and separation.
The encoder converts $\mathbf{X}$ from the sensor into high-dimensional representations $\mathbf{Y} \in \mathbb{R}^{256 \times \frac{m}{16}}$ to provide more detailed information.
Then abstraction retrieves the features  $\mathbf{Z} \in \mathbb{R}^{2048 \times \frac{m}{256}}$ related to anomalies, and finally the separation pairs $\mathbf{Z}$ to each $i$-th machine $\mathbf{f}_i \in \mathbb{R}^{1 \times 256}$.
Based on these anomaly features, the anomaly score can be obtained directly.

\subsection{In-Network (Pre-)Processing}

As a \ac{nn} model, the processing logic of information abstraction is often deployed centralized on a dedicated server/cloud due to the large computational requirements.
However, with the centralized deployment, all raw sensory data $\mathbf{X}$ needs to be transmitted through the network to this server/cloud, i.e., the entire sequence of the generated audio of each client.
Such a large volume of data will inevitably lead to network congestion and data transmission delays.

An alternative way is to deploy the processing logic on an edge cloud near the individual \ac{iot} sensors in order to achieve shorter transmission delays.
However, due to the nature of networks, the computational resources (in this case, processing and memory) are limited on edge nodes.
This limitation increases the processing delay, which in effect results in limited or no reduction in the total service latency.
As an example, $9.08 \times 10^9$ operations are needed by \ac{ia-net}~\cite{Wu2106:Abstraction} to complete the processing task.

Fortunately, information abstraction can eliminate redundant information.
With the development of network softwarization, computational resources in the network become feasible for performing intermediate processing.
Therefore, we can consider that the entire processing logic is partitioned into several \acp{vnf}
that filter redundant data one by one.
When these \ac{vnf}s consume only a small amount of computational resources, the resource-constrained network nodes can perform partial data processing tasks without adding additional processing latency.
At the same time, through these \ac{vnf}s, redundant data can be gradually filtered out without causing network congestion and increased transmission latency.
The nodes in the network also collaborate toward optimizing the overall service latency.

Therefore, following the above idea, this paper takes industrial anomaly detection as a case study to solve the following two problems:
\begin{enumerate}
    \item How to design the processing logic for the resource-constrained network nodes so that it is deployable as a \ac{vnf} without extra processing delay?
    \item How should such a \ac{vnf} be deployed in the network to achieve the purpose of redundant data filtering?
\end{enumerate}

\section{Our Solution: IA-Net-Lite}\label{sec:method}

\begin{figure}[t!]
\centerline{\includegraphics[width=.9\columnwidth]{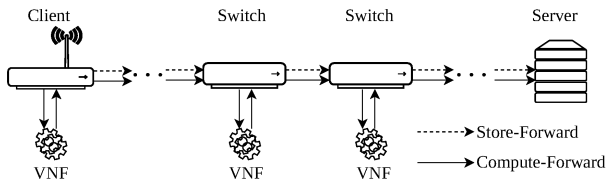}}
\caption{Overview of the in-network data processing scenario.}
\label{fig:ia-net-lite-emu}
\end{figure}

In this section, we first show the design of our lightweight \ac{nn} model-based \acrfull{ia-net-lite}, so that it is suitable for deployment on resource-constrained network nodes in order to abstract information.
Secondly, we discuss how it should be partitioned and deployed to filter traffic to optimize service latency.

\cref{fig:ia-net-lite-emu} illustrates the scenario of in-network data processing.
The collected data is constantly sent to a network client, where the client can be either a WiFi access point or a cellular base station of a (non-public) network. This client keeps forwarding the data to the backend ``server'' over a forwarding path.
This path can be either dynamically determined based on a routing mechanism or statically configured.
The forwarding path consists of some intermediate network switches, which support \ac{sdn} and \ac{nfv}.

In a traditional communication network, the network nodes (i.e., clients and switches) are only responsible for forwarding data to the server, and the entire \ac{nn} model is deployed on the server to process the forwarded data.
This approach is usually called \ac{sf}.
In the scenario of the in-network process, the \ac{nn} model is split into several sub-operations, which are deployed as \ac{vnf}s on the network nodes along the data path.
These network nodes perform more than just forwarding data: they first process the data using the \ac{vnf} and then forward the intermediate results to the next node.
This is commonly known as \ac{cf}.

\subsection{A Lightweight \acrlong{nn} Model for Networks}\label{sec:lightweight_design}


As the resources (computing and storage) of network devices are limited, the overall parameters and operations of a \ac{nn} model should be tightly controlled at an acceptable scale, which is suitable for such lightweight devices.
The complexity of a \ac{nn} model during inference is defined by the memory and computing power required to process the input data. This can be mathematically formulated by the number of the parameters (Param) and the number of the \acp{mac}.

To minimize Param and \ac{mac}, we designed \ac{ia-net-lite} for in-network processing, inspired by \ac{ia-net} in~\cite{Wu2106:Abstraction}.
Under the guidance of the paradigm from \ac{ia-net}, \ac{ia-net-lite} also consists of three distinct modules: encoder, abstraction, and separation.

\subsubsection{Encoder}
In the encoder module, the one-dimensional input data is transformed into a higher dimensional representation, in order to provide more detailed information for subsequent operations.
IA-Net-Lite down-samples and maps the mixed input data $\mathbf{X} \in \mathbb{R}^{1 \times m}$ in a high-dimensional STFT-like feature space $\mathbf{Y} \in \mathcal{R}^{N \times \frac{m}{4}}$ by using a 1-D convolutional layer (1D-Conv) with stride four.
The 1D-Conv is a set of filters consisting of $N$ filters with length $L$.
In the encoder module, only one convolutional layer is used to minimize computation and storage resources.
This module eventually only contains $0.18 \times 10^3$ Params, and $3.00 \times 10^6$ \ac{mac}s.

\subsubsection{Abstraction}
\begin{figure}[t!]
\centerline{\includegraphics[width=.75\columnwidth]{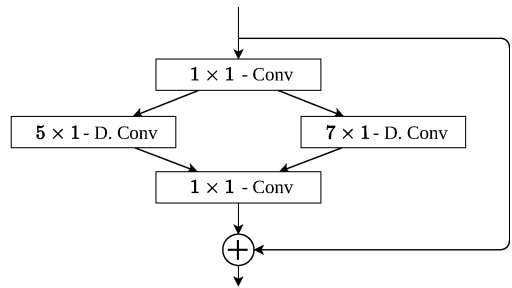}}
\caption{1D-residual convolution (1D-R-Conv).}
\label{fig:d_conv}
\end{figure}
The objective of this module is to extract anomaly information $\mathbf{F}$ from the encoded features $\mathbf{Y}$.
The abstract module accounts for most of the parameters, which is the main part of the model to be compressed.
Therefore, we propose a 1D-residual convolution (1D-R-Conv).
As illustrated in~\cref{fig:d_conv}, it follows three strategies: ($i$) depth-wise residual block~\cite{sandler2018mobilenetv2} to reduce the computation; ($ii$) a large down-sampling rate of four to squeeze the input data; ($iii$) a dual path~\cite{9367428} structure to obtain enriched semantic and temporal domain information and to obtain a balance between accuracy and size.

Empirically, the commonly used $3\times1$ convolution kernel causes information loss when the down-sampling rate is four, e.g., ResNet~\cite{he2016deep}.
To avoid information loss during information abstraction while increasing the receptive field of the convolutional layer, we use a large convolutional kernel $7 \times 1$ and $5 \times1$ in the dual path structure.
Each convolution operation is followed by a non-linearity layer (ReLU) and a layer normalization (Norm).
The abstraction module consists of several stacked 1D-R-Conv blocks.
Although the abstraction module is most resource-intensive, with the lightweight design it only requires $2.29 \times 10^6$ Params and $416.00 \times 10^6$ \ac{mac}s.
\subsubsection{Decoder}
In the decoder module we utilize an average pooling layer (Avg.) to squeeze the temporal information to $1$, and replace the fully-connected layer in the original IA-Net with a $1 \times 1$ convolutional layer (1D-Conv) to assign the anomalous representations to $n$ objects as abstracted features $\mathbf{F} \in \mathbb{R}^{n \times 256}$.
By reducing the number of parameters, the decoder module has $1.31 \times 10^6$ Params and $420.00 \times 10^6$ \ac{mac}s, which is more conducive to deployment in the network.

\subsection{Deployment in the Network}\label{sec:offloading}
\begin{table}[t!]
    \vspace{2mm}
    \def\arraystretch{1.5}
    \centering
    \caption{The traffic filter of each operations block in IA-Net-Lite.}
    \label{tab:layers_ia_net_lite}
    \begin{tabular}{C{2.1cm}|C{2cm}|C{1.5cm}|C{1.5cm}}
    \toprule[1pt]
    \bf Operation Block & \bf Module & \bf Input & \bf Output\\
    \midrule[1pt]
    Encoder & 1D-Conv           & $1 \times m$ & $32 \times \frac{m}{4}$\\
    Layer 1 & 1D-R-Conv$\times$1  & $32 \times \frac{m}{4}$ & $16 \times \frac{m}{4}$\\
    Layer 2 & 1D-R-Conv$\times$2  & $16 \times \frac{m}{4}$ & $24 \times \frac{m}{16}$\\
    Layer 3 & 1D-R-Conv$\times$3  & $24 \times \frac{m}{16}$ & $32 \times \frac{m}{64}$\\
    Layer 4 & 1D-R-Conv$\times$4  & $32 \times \frac{m}{64}$ & $64\times\frac{m}{256}$\\
    Layer 5 & 1D-R-Conv$\times$3  & $64\times\frac{m}{256}$ & $96\times\frac{m}{256}$\\
    Layer 6 & 1D-R-Conv$\times$3  & $96\times\frac{m}{256}$ & $160\times\frac{m}{1024}$\\
    Layer 7 & 1D-R-Conv$\times$1  & $160\times\frac{m}{1024}$ & $320\times\frac{m}{1024}$\\
    Layer 8 & 1D-Conv           & $320\times\frac{m}{1024}$ & $1280\times\frac{m}{1024}$\\
    Separation & Avg.1D-Conv    & $1280\times\frac{m}{1024}$ & $4\times 256$\\
    \bottomrule[1pt]
    \end{tabular}
\end{table}
In a network, usually more data is transmitted than is needed or expected.
Therefore, one solution for reducing the amount of data is to use an information filtering logic as a \ac{vnf} that can be deployed on the network nodes to filter out redundant or unwanted data before the traffic leaves each node.

The limitations on the available resources (processing and memory) of the network nodes is addressed by the lightweight designed \ac{nn} model \ac{ia-net-lite} in~\cref{sec:lightweight_design}.
Therefore, compute resources inside the network are available and intermediate processing can be performed in the network and the operation blocks of \ac{ia-net-lite} can be deployed as \ac{vnf} within the network.
However, the question of how computing and networking are integrated, i.e., which operation blocks the deployed \ac{vnf}s should be composed of, is intuitively important.
A proper design will reduce the data volume at each hop, consequently, preventing network congestion, and, ultimately, improving the overall service latency.

In~\cref{tab:layers_ia_net_lite}, the different operation block outputs in relation to the input data for \ac{ia-net-lite} is listed.
To better assess the ability of traffic filtering of the different operation blocks, we define filter rate $r$ as follows:
\begin{equation}
    r = \frac{\text{Size of Operation Block Output}}{\text{Size of System Input}},
\end{equation}
which indicates how much data has been filtered out by the operation block.
An $r > 1$ means that the input data is augmented, and vice versa.

\begin{figure}[t!]
\centerline{\includegraphics[width=\columnwidth]{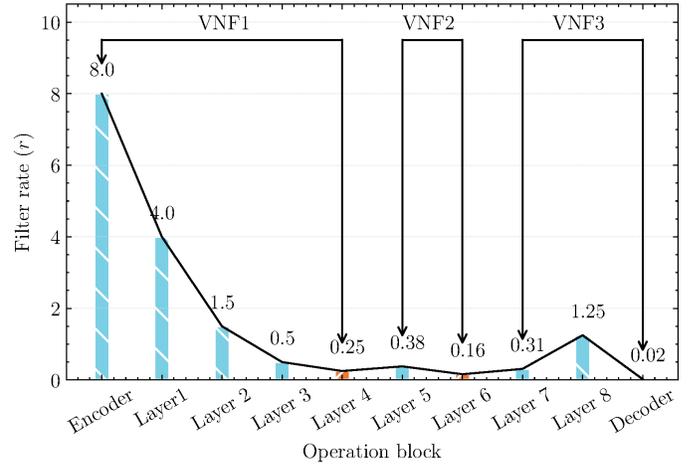}}
\caption{Filter rate $r$ of data flows by each operation block and the splitting of the \acp{vnf}.}
\label{fig:layers_compress}
\end{figure}
The trend of traffic filtering efficiency is fitted to the $r$ of all operation blocks together.
To maximize the filtering efficiency on every link in the network, we select the concave points in this filtering trend as the bounds for constructing the \ac{vnf}s.
The concave point of the $r$-curve is represented by the last operation of the continuous data reduction.
Splitting before this concave point does not maximize the filtering effect, however, splitting after this point will cause the filtering effect to deteriorate.

As shown in~\cref{fig:layers_compress}, the $r$ of Layer 4 and Layer 6 are two concave points, which split \ac{ia-net-lite} into three \ac{vnf}s.
By these three \ac{vnf}s, the data volume can be reduced to $25\%$, $16\%$, and $2\%$ of the original input data, respectively.
Therefore, the overall service latency can be improved when these three \ac{vnf}s are deployed on network nodes.
In~\cref{sec:service_latency} we show the evaluation of the effectiveness of traffic filtering.

\section{Experimental Design}\label{sec:experimental_design}
In this section, we describe our experimental design, including the emulation setups and measurement metrics.

\subsection{Emulation Setups}\label{subsec:evaluation_setup}
In our evaluation, we use the realistic dataset of MIMII~\cite{Purohit_DCASE2019_01}, which is a widely used acoustic data set for malfunctioning industrial machine inspection, containing 26092 normal and anomalous operating acoustic segments from four types of real machinery: valves, pumps, fans, and slide rails.
Each acoustic segment is a single-channel 10-second-long acoustic segment with a sampling rate of 16000~Hz.
We randomly select all $n=4$ types of sources from the dataset to construct the source signals $\mathbf{S}$.
We mixed the source signals $\mathbf{S}$ following the standard normal distribution to simulate as many mixing scenarios as possible.
By doing this, our evaluation has covered various data types and mixing cases.

We conducted the evaluation on the network emulator \ac{comnetsemu}~\cite{xiang2020comnetsemu}.
As illustrated in~\cref{fig:ia-net-lite-emu}, we set up a multi-hop topology with switches as intermediate network nodes.
Each node can perform either one or both operations of forwarding and data processing.
Hence, there are two different modes for the deployment of the anomaly detection \ac{nn} models, which are \acrfull{cf} and \acrfull{sf}.

We assume that the clients connect to the network and send the observed data using User Datagram Protocol (UDP) to the server.
All topology links have the same homogeneous bandwidth of $10$~Mbps and a fixed propagation delay of $150$~ms.
In the emulation, we do not consider packet losses and assume that all packets are received sequentially.
These settings can be considered typical for an edge network. We also assume that all devices and connections are general-purpose and available commercially as is.
We implement the client and server functions in Python.
For each configuration of a parameter set, we performed 60 measurements in which the client sends the observed data to the server, to ensure the experimental results being statistically significant.

Each packet is forwarded directly to the server by the network node at the fastest possible \ac{sf} speed in conventional \ac{sf} mode.
When operating in the \ac{cf} mode, the \ac{sdn} controller inserts rules into the flow table of each switch to forward the data traffic through each middlebox to build an \ac{sfc}.
For the overall experiments, we employ a \ac{cots} server with a 2.0G Intel i5 CPU with 16GB RAM using Ubuntu 18.04 LTS.

\subsection{Metrics}\label{subsec:evaluation_metrics}

For the evaluation of our method we employ the following metrics:

\subsubsection{Service Latency}
Service latency $t_s$ is the \ac{rtt} between the request sent by the client and the response delivered by the server.
It includes all transmission and computation delays introduced by end nodes (client and server) and network nodes.
$t_s$ greatly affects the QoS of network services.
We compare $t_s$ of \ac{ia-net-lite} in \ac{cf} and \ac{sf} modes, and \ac{ia-net} in \ac{sf} mode.

\subsubsection{Computation Load}
To measure the computation load of an \ac{vnf}, we introduce the metric of processing latency $t_p$, which is the computation time required by a device to complete the \ac{vnf} deployed on it.
With the same computational power, a smaller $t_p$ means that the \ac{vnf} on the device is completed faster.
As a result, the computation load of this \ac{vnf} is small.
Conversely, a larger $t_p$ implies a large computational load for this task.

\subsubsection{Transmission Latency}
Transmission latency $t_t$ is the time taken by the network to transmit the packets over the links, including propagation delay, queuing delay, and transmission delay.
For the same network setup -- e.g., bandwidth, mac access, and so on -- $t_t$ is closely related to the size of data transmitted on the links.

Overall, in our experimental setups one can assume that $t_s = t_p + t_t$.
\section{Results and Discussions}\label{sec:results_analysis}
Based on practical experimental results from emulation, we evaluate and discuss the performance of the in-network processing method for low-latency industrial anomaly detection.

\subsection{Service Latency}\label{sec:service_latency}
\begin{figure}[t!]
\centerline{\includegraphics[width=\columnwidth]{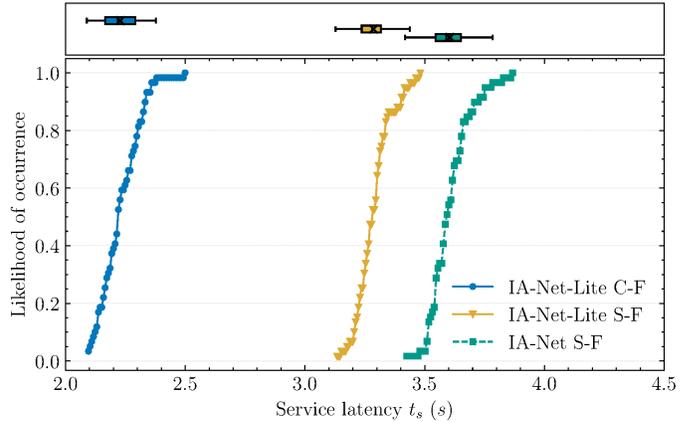}}
\caption{Likelihood of end-to-end service latency in \ac{cf} and \ac{sf} modes.}
\label{fig:service_latency_likelihood}
\end{figure}

\cref{fig:service_latency_likelihood} shows the service latency of \ac{ia-net-lite} in \ac{cf} and \ac{sf} modes, and \ac{ia-net} in \ac{sf} mode, respectively.
We observed that in \ac{sf} mode, the median service latency is reduced from $3.6$s in \ac{ia-net} to $3.3$s in \ac{ia-net-lite}, which is a  $8.33\%$ reduction.
This decrease is due to the fact that our lightweight design makes \ac{ia-net-lite} consume much less computational resources, resulting in less processing latency on the server.
\cref{sec:processing_latency} discusses the lightweight \ac{nn} in detail.

More importantly, comparing the median service latency of \ac{ia-net-lite} in \ac{cf} and \ac{sf} mode, this latency is significantly reduced from $3.3$s to $2.2$s.
This is due to the collaboration of network nodes when filtering out redundant information and thus reducing the transmission latency.
In other words, the service latency is reduced by $33.33\%$ in the in-network processing scheme.
Compared to the baseline system \ac{ia-net}, this latency is significantly reduced by $38.89\%$.
We discuss traffic filtering in the network in more detail in~\cref{sec:transmission_latency}.

To summarize, we propose \ac{ia-net-lite} in this paper for in-network processing, which can significantly decrease the service latency as it is capable of reducing both the transmission and processing latency with the help of network nodes.

\subsection{Processing Latency and Lightweight Design}\label{sec:processing_latency}
\begin{figure}[t!]
\centerline{\includegraphics[width=\columnwidth]{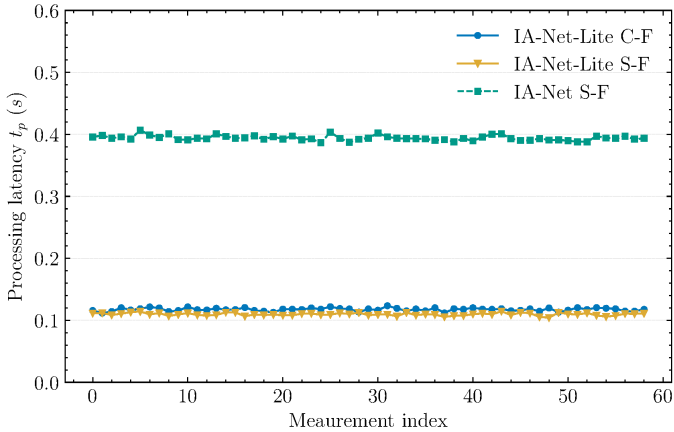}}
\caption{Overall processing latency in \ac{cf} and \ac{sf} modes.}
\label{fig:processing_latency}
\end{figure}


\cref{fig:processing_latency} shows the processing latency of the proposed \ac{ia-net-lite} and the baseline system \ac{ia-net}, to verify the lightweight design of \ac{ia-net-lite}.
In \ac{sf} mode, we first observe that the processing latency of \ac{ia-net-lite} is about $0.1$s, which is a mere quarter of \ac{ia-net}'s.
On the one hand, the reduction of the overall processing latency explains the gap between \ac{ia-net-lite} and \ac{ia-net} with the \ac{sf} mode in~\cref{fig:service_latency_likelihood}.
Moreover, while the server remains the same, reducing the processing latency by quarter means that the computation load required by \ac{ia-net-lite} is reduced to a quarter.
Thus, \ac{ia-net-lite} with the lightweight design is more suitable for deployment on resource-constrained network nodes.

Moreover, we observe that \ac{ia-net-lite}'s processing latency in \ac{sf} and \ac{cf} are different. In \ac{cf}, it is slight about $0.005$s ($5\%$) more than that in \ac{sf}.
Since \ac{ia-net-lite} uses exactly the same processing logic in both \ac{sf} and \ac{cf} modes, one of the reasons why this happens could be due to the I/O processing delay between the switch and \ac{vnf} in \ac{cf}.
However, we consider the instability of the processing time to be within the acceptable range compared to the benefit of reduced service latency.

\subsection{Transmission Latency and Traffic Filter}\label{sec:transmission_latency}

\begin{table}[t!]
    \vspace{2mm}
    \def\arraystretch{1.5}
    \centering
    \caption{Data throughput and filter rate $r$ on each link, and the total transmission latency $t_t$ of \ac{ia-net-lite} in \ac{cf} and \ac{sf} modes.}\label{tab:transmission_latency}
    \begin{tabular}{C{1cm}|C{1.5cm}|C{1.5cm}|C{1.5cm}|C{1cm}}
    \toprule[1pt]
    \bf Mode  & \bf client$\rightarrow$s1 & \bf s1$\rightarrow$s2 & \bf s2$\rightarrow$server & $\mathbf{t_t}$\\
    \midrule[1pt]
    \multirow{2}{*}{\shortstack{\ac{cf}}}
    & 64397 byte & 40715 byte & 17626 byte & \multirow{2}{*}{\shortstack{$2.1s$}}\\
    \cline{2-4}
    & $25.12\%$ & $15.88\%$ & $6.87\%$ \\
    \hline
    \multirow{2}{*}{\shortstack{\ac{sf}}}
    & 256406 byte & 256406 byte & 256406 byte & \multirow{2}{*}{\shortstack{$3.2s$}}\\
    \cline{2-4}
    & $100\%$ & $100\%$ & $100\%$ \\
    \bottomrule[1pt]
    \end{tabular}
\end{table}

To demonstrate the effectiveness of the traffic filter in terms of transmission latency, we listed the data throughput and the filter rate on each network connection in~\cref{tab:transmission_latency}, and gave the results in \ac{sf} mode for comparison.
Since all bandwidths between network nodes in our emulator setup are $10$~Mbps with $150$~ms delay, reducing data throughput effectively leads to a decrease in transmission latency.

\cref{tab:transmission_latency} shows that after the filtering of \ac{vnf}s on the client and the switch 1 (s1),
the bandwidth between them is consuming only $25.12\%$ and $15.88\%$ of the original data, respectively, while the high-density information reaching the server is compressed down to $6.87\%$ of the original data after the last \ac{vnf} on the switch 2 (s2).
Therefore, the total transmission latency $t_t$ in \ac{cf} mode is reduced to about $2.1$s.
This reduction, expressed in the overall service latency, is the gap between \ac{ia-net-lite} with the \ac{sf} and \ac{cf} mode in~\cref{fig:service_latency_likelihood}.

In contrast, the bandwidth consumption of the traditional \ac{sf} remains the same as the original data, and thus the transmission latency cannot be reduced.
$t_t$ stays about $3.2$s with \ac{sf} mode.
Comparing the \ac{cf} and \ac{sf} modes, although they hold exactly the same processing latency, the redundant data is filtered by the \ac{vnf}s on the network nodes with the help of the in-network processing scheme, thus reducing the transmission latency and finally obtaining a smaller service latency.

\section{Conclusion}\label{sec:conclusion}

Time-critical applications are becoming more and more important in modern smart factories, however, these applications have delayed data transmission and processing due to high-density sensors and large data volumes. In this paper, we show our brand new design for an in-network data processing with the help of network softwarization for industrial anomaly detection, which we call \acrfull{ia-net-lite}.
This approach leverages for the first time the resources of increasingly intelligent network devices to combine data transmission and processing with filters that reduce redundancy, and, ultimately, optimize service latency. 
In this paper, we showed via the employment of a practical network emulator that our proposed approach can reduce the service latency by up to 38.89\%.
The design idea of this method is also applicable to other large-volume and AI applications.

However, comparing with the theoretical filter rate $r$ in~\cref{sec:offloading}, the measured $r$ (see~\cref{tab:transmission_latency}) in our emulations is slightly larger.
This is due to that the \ac{cf} can only filter the payloads of the network packets, while the headers remain unchanged.
Header compression techniques~\cite{Tomo1905:Unidirectional} can be used to reduce the packet headers, thereby maximizing the filtering effect.
This is in our future research agenda.


\section*{Acknowledgment}
This work was funded by the Federal Ministry of Education and Research of Germany (Software Campus Net-BliSS 01IS17044), and by the German Research Foundation (Deutsche Forschungsgemeinschaft) as part of Germany’s Excellence Strategy (EXC 2050/1 – Project ID 390696704) Cluster of Excellence “Centre for Tactile Internet with Human-in-the-Loop” (CeTI) of Technische Universität Dresden.

\bibliographystyle{IEEEtran}
\bibliography{bibtex}

\end{document}